\documentclass[a4 paper, 12 pt] {article}

\begin{document}
\title{\bf{Toroidal Soliton Solutions in $O(3)^N$ Nonlinear Sigma Model.}}
\author{A. Wereszczy\'{n}ski $^{a)}$ \thanks{wereszcz@alphas.if.uj.edu.pl}
       \\
       \\ $^{a)}$ Institute of Physics,  Jagiellonian University,
       \\ Reymonta 4, Krak\'{o}w, Poland}
\maketitle
\begin{abstract}
A set of N three component unit scalar fields in $(3+1)$ Minkowski
space-time is investigated. The highly nonlinear coupling between
them is chosen to omit the scaling instabilities. The
multi-soliton static configurations with arbitrary Hopf numbers
are found. Moreover, the generalized version of the
Vakulenko-Kapitansky inequality is obtained. The possibility of
attractive as well as repulsive interaction between hopfions is
shown. A noninteracting limit is also discussed.
\end{abstract}
keyword: Hopf solitons
\\
\newpage
%%%%%%%%%%%%%%%%%%%%%%%%%%%%%%%%%%%%%%%%%%%%%%%%%%%%%%%%%%%%%%%
\section{\bf{ Introduction}}
%%%%%%%%%%%%%%%%%%%%%%%%%%%%%%%%%%%%%%%%%%%%%%%%%%%%%%%%%%%%%%%
Topological defects are one of the main ideas of the temporary
theoretical physics. They play crucial role in many models - from
cosmological strings and textures \cite{vilenkin}, \cite{davis} to
magnetic monopoles and vortices in Quantum Chromodynamics
\cite{wu}, \cite{thooft}. On the other hand, topological defects
are observed in various condense matter experiments - see for
instance liquid crystals and $^3He$ \cite{volovik} or $^4He$
quantum liquids \cite{donnelly}. Among quite well understood
topological defects like domain walls, strings/vortices and
monopoles toroidal-like configurations, that is solitons with
nonzero value of the Hopf index, seem to be rather mysterious
objects.
\\
Since Faddeev, Niemi \cite{niemi}, \cite{langmann} and Cho
\cite{cho} have proposed their famous effective model for the low
energy gluodynamics, where glueballs (particles which consist of
only the gauge fields) are knot-like objects made of self-linking
flux-tubes, toroidal solitons, their physical as well as
mathematical aspects have been considered by many theoreticians
\footnote[1]{Knotted solitons appear also in condensed matter
\cite{babaev1} and possibly in astrophysics \cite{babaev2}.}.
Unfortunately, due to the rather complicated toroidal symmetry and
nonlinearity of equations of motion no analytical results, in
recently investigated QCD relevant models \cite{niemi2},
\cite{sanchez1}, \cite{my1}, have been found. Almost all results
have been obtained by means of numerical methods \cite{battyde},
\cite{salo}, \cite{ward}, \cite{wipf}. In order to deal with exact
toroidal solutions one is forced to consider various toy models.
\\
All such theories, providing analytical description of hopfions,
are based on the quite old idea presented by Deser et al.
\cite{deser}. They have investigated a classical highly nonlinear
field theory with a form of the Lagrangian chosen to circumvent
Derrick's theorem. In fact a soliton solution with unit Hopf index
\cite{nicole} and then configurations with arbitrary value of the
Hopf invariant have been found \cite{aratyn}. Moreover, nontrivial
topological solutions have been obtained in more complicated model
\cite{sanchez2}, for example with explicitly broken $O(3)$
symmetry \cite{my2}. Using the exact soliton solutions one has
checked that the Vakulenko-Kapitanski \cite{vakulenko} inequality
is fulfilled in all these models.
\\
It should be mentioned that soliton solutions with non-trivial
Hopf index can be achieved also in another famous model i.e.
Skyrme model in $(3+1)$ dimensions \cite{stern}, \cite{cho_sk}.
Moreover, there is also possibility to analyze hopfions in less
dimensional space-time. It is based on the observation that the
Faddeev-Niemi knots can be understood as twisted magnetic vortex
rings built of so-called new baby skyrmions found in $(2+1)$
anisotropic Skyrme model \cite{skyrme}.
\\
In the present paper we would like to investigate toroidal soliton
solutions in a model which consists of many unit vector fields
i.e. allows for many topological charges. The appearance of more
than one unit vector field has been observed in recently derived
non-Abelian generalization of the color dielectric model
\cite{my1}. Indeed, there are two unit fields in this model. The
main aim, besides finding exact form of the topological solutions
and their energies and Hopf charges, is to analyze
Vakulenko-Kapitanski formula in case of more than one Hopf index.
%%%%%%%%%%%%%%%%%%%%%%%%%%%%%%%%%%%%%%%%%%%%%%%%%%%%%%%%%%%%%%%
\section{\bf{ The model}}
%%%%%%%%%%%%%%%%%%%%%%%%%%%%%%%%%%%%%%%%%%%%%%%%%%%%%%%%%%%%%%%
Let us consider a set of $N$ unit three component scalar fields
$\vec{n}_i, \, i=1...N$, $\vec{n}_i^2=1$, living in the $(3+1)$
Minkowski space-time. The Lorentz and $O(3)^N$ invariant
Lagrangian density is chosen in the following form
\begin{equation}
\mathcal{L}=\prod_{i=1}^{N} \left( [\vec{n}_i \cdot
(\partial_{\mu} \vec{n}_i \times \partial_{\nu} \vec{n}_i
)]^2\right)^{\alpha_i} \label{model1}
\end{equation}
which is a generalized version of the Aratyn-Fereira-Zimerman
model with $N$ unit fields. It can be rewritten as
$\mathcal{L}=\prod_{i=1}^N (H_{\mu \nu}^{(i)2})^{\alpha_i}$, where
$H_{\mu \nu}^{(i)} =\vec{n}_i \cdot ( \partial_{\mu } \vec{n}_i
\times
\partial_{\nu } \vec{n}_i )$ are antisymmetric field tensors.
In order to omit the Derick arguments for nonexistence of static
soliton solutions one has to assume the condition
\begin{equation}
\sum_{i=1}^N \alpha_i = \frac{3}{4}. \label{stab_cond}
\end{equation}
In fact, now the total energy becomes invariant under the scale
transformations. \\
Now, after taking advantage of the stereographic projection, we
can express each of the unit fields in terms of two complex
functions $u$ and $u^*$:
\begin{equation}
\vec{n}_i= \frac{1}{1+|u_i|^2} ( u_i+u^*_i, -i(u_i-u^*_i),
|u_i|^2-1). \label{stereograf}
\end{equation}
Then the Lagrangian density reads
\begin{equation}
\mathcal{L}=8^{3/4} \prod_{i=1}^{N}
\frac{1}{(1+|u_i|^2)^{4\alpha_i}} (K_{\mu}^{(i)}
\partial^{\mu} u^*_i)^{\alpha_i}, \label{model2}
\end{equation}
where we have introduced a set of the objects:
\begin{equation}
K_{\mu}^{(i)} =(\partial_{\nu} u^*_i \partial^{\nu} u_i)
\partial_{\mu} u_i -(\partial_{\nu}u_i \partial^{\nu } u_i)
\partial_{\mu}u^*_i, \label{K_i}
\end{equation}
where $i=1...N$. It is straightforward to check that they fulfill
the following conditions
\begin{equation}
K_{\mu}^{(i)} \partial^{\mu} u_i=0 \; \; \; \; \; \mbox{and} \; \;
\; \; \; Im(K_{\mu}^{(i)}
\partial^{\mu} u^*_i )=0. \label{properties}
\end{equation}
The equations of motion take the form
$$
\partial_{\mu} \left[ \left( \prod_{i=1, i \neq j}^{N}
\frac{(K_{\mu}^{(i)}
\partial^{\mu} u^*_i)^{\alpha_i}}{(1+|u_i|^2)^{4\alpha_i}} \right)
\frac{2 \alpha_j (K_{\mu}^{(j)}
\partial^{\mu} u^*_j)^{\alpha_j-1} }{(1+|u_j|^2)^{4\alpha_j}} K^{(j)
\mu}\right] -
$$
\begin{equation}
- \prod_{i=1, i \neq j}^{N} \frac{(K_{\mu}^{(i)}
\partial^{\mu} u^*_i)^{\alpha_i}}{(1+|u_i|^2)^{4\alpha_i}}
(K_{\mu}^{(j)}
\partial^{\mu} u^*_j)^{\alpha_j} \left[
\frac{1}{(1+|u_j|^2)^{4\alpha_j}} \right]_{u_j^*}' =0.
\label{eq_mot1}
\end{equation}
After some calculation one can rewrite them as follow
\begin{equation}
\partial_{\mu} \left[ \left( \prod_{i=1, i \neq j}^{N}
\frac{(K_{\mu}^{(i)}
\partial^{\mu} u^*_i)^{\alpha_i}}{(1+|u_i|^2)^{4\alpha_i}} \right)
\frac{(K_{\mu}^{(j)}
\partial^{\mu} u^*_j)^{\alpha_j-1} }{(1+|u_j|^2)^{4\alpha_j -2}} K^{(j)
\mu}\right] =0,  \label{eq_mot2}
\end{equation}
or in the more compact form
\begin{equation}
\partial_{\mu} \mathcal{K}^{(j) \mu}=0, \label{eq_mot3}
\end{equation}
where
\begin{equation}
\mathcal{K}_{\mu}^{(j)}= \left( \prod_{i=1, i \neq j}^{N}
\frac{(K_{\mu}^{(i)}
\partial^{\mu} u^*_i)^{\alpha_i}}{(1+|u_i|^2)^{4\alpha_i}} \right)
\frac{(K_{\mu}^{(j)}
\partial^{\mu} u^*_j)^{\alpha_j-1} }{(1+|u_j|^2)^{4\alpha_j -2}}
K^{(j) \mu}. \label{K_def}
\end{equation}
It is easy to show that these quantities also fulfill the
previously found conditions (\ref{properties}). Because of that we
can generalize the procedure defined in \cite{aratyn} and
construct the infinity families of the conserved currents
\begin{equation}
\mathcal{J}_{\mu}^{(i)} = \mathcal{K}^{(i) \mu} \frac{ \partial
G_i}{\partial u_i}- \mathcal{K}^{* (i) \mu } \frac{ \partial
G_i}{\partial u^*_i}, \label{current}
\end{equation}
where $G_i, \; i=1...N$ are any functions of $u_j$ and $u^*_j$,
$j=1..N$. Due to that presented $O(3)^N$ invariant model is
integrable in the sense that $N$ infinite families of the
conserved currents can be found \cite{alvarez}, \cite{ferreira}.
\\
Let us proceed further and prove that, as in case of soliton
models in $(1+1)$ and $(2+1)$ dimension, the existence of infinite
number of the conserved currents leads to topological soliton
solutions. We begin with introducing of the toroidal variables
$$ x=\frac{a}{q} \sinh \eta \cos \phi , $$
$$ y=\frac{a}{q} \sinh \eta \sin \phi , $$
\begin{equation}
z=\frac{a}{q} \sin \xi ,\label{tor_coord}
\end{equation}
where $q= \cosh \eta -\cos \xi$ and the constant $a>0$ sets the
scale of the coordinates. However, one should keep in mind that
the model is scale invariant. Due to that, there is no scale on
the solution level. All solitons, irrespective of size, have
identical energy.
\\ In addition we assume a natural generalization of the
Aratyn-Feriera-Zimer\-man Ansatz \cite{aratyn}
\begin{equation}
u_i(\eta,\xi,\phi) \equiv f_i(\eta) e^{i(m_i\xi + n_i \phi)}, \;
i=1...N, \label{anzatz}
\end{equation}
where $m_i, n_i, \;  i=1...N$ are integral parameters.\\
Inserting the Ansatz into the field equations (\ref{eq_mot3}) we
derive at the following static equations
$$
\partial_{\eta} \ln \left[ \left( \prod_{i=1, i\neq j}^N
\frac{(f_if'_i)^{2\alpha_i}}{(1+f_i^2)^{4\alpha_i}} \right)
\frac{(f_jf'_j)^{2\alpha_j -1}}{(1+f_j^2)^{4\alpha_j -2}}\right] =
$$
\begin{equation}
= -
\partial_{\eta} \ln \left[ \sinh \eta \prod_{i=1}^N
\left( m_i^2 +\frac{n_i^2}{\sinh^2 \eta}\right)^{\alpha_i}
\right]. \label{eq_mot4}
\end{equation}
These equations can be integrate and we obtain that
\begin{equation}
\left( \prod_{i=1, i\neq j}^N
\frac{(f_if'_i)^{2\alpha_i}}{(1+f_i^2)^{4\alpha_i}} \right)
\frac{(f_jf'_j)^{2\alpha_j -1}}{(1+f_j^2)^{4\alpha_j -2}} =
\frac{k_j}{\sinh \eta} \prod_{i=1}^N \left( m_i^2
+\frac{n_i^2}{\sinh^2 \eta}\right)^{-\alpha_i}, \label{eq_mot5}
\end{equation}
where the integration constants $k_j, \; j=1...N$ have been
introduced. \\
One can show that (\ref{eq_mot5}) can be reduced to the set of $N$
decoupled first order differential equations
\begin{equation}
\frac{f_jf'_j}{(1+f_j^2)^2}= \frac{1}{k_j} \left( \prod_{i=1}^N
k_i^{4\alpha_i} \left( m^2_i +\frac{n^2_i}{\sinh^2 \eta
}\right)^{-2\alpha_i} \right) \frac{1}{\sinh^2 \eta},
\label{eq_mot6}
\end{equation}
with the following general solutions
\begin{equation}
\frac{1}{1+f^2_j} = \frac{1}{k_j} \int \left( \prod_{i=1}^N
k_i^{4\alpha_i} \left( m^2_i +\frac{n^2_i}{\sinh^2 \eta
}\right)^{-2\alpha_i} \right) \frac{d\eta}{\sinh^2 \eta} + l_j.
\label{sol}
\end{equation}
Here $l_j, \; j=1...N$ are integration constants. \\
Now, we will analyze the total energy corresponding to our model
(\ref{model1}):
\begin{equation}
E \equiv \int d^3x T_{00} = 8^{3/4} \int d^3x \prod_{i=1}^{N}
\frac{(K_{j}^{(i)}
\partial^{j} u^*_i)^{\alpha_i}}{(1+|u_i|^2)^{4\alpha_i}}, \label{energy1}
\end{equation}
where the stereographic projection (\ref{stereograf}) has been
taken into account. Inserting (\ref{K_i}) and the Ansatz
(\ref{anzatz}) into (\ref{energy1}) we derive that
\begin{equation}
E_{m,n}=(2\pi)^2 8 \cdot 2^{3/4} \int_0^{\infty} d \eta \sinh \eta
\left( \prod_{i=1 }^N
\frac{(f_if'_i)^{2\alpha_i}}{(1+f_i^2)^{4\alpha_i}} \left( m_i^2
+\frac{n_i^2}{\sinh^2 \eta}\right)^{\alpha_i}
\right)\label{energy2}
\end{equation}
Quite surprisingly, using the equations (\ref{eq_mot6}) we are
able to remove the unknown functions $f_j$ from this formula. Then
the total energy integral takes the form
\begin{equation}
E_{m,n}=(2\pi)^2 8 \cdot 2^{3/4} \prod_{i=1}^N k_i^{4\alpha_i}
\int_0^{\infty} \frac{d \eta }{\sinh^2 \eta} \prod_{i=1 }^N \left(
m_i^2 +\frac{n_i^2}{\sinh^2 \eta}\right)^{-2\alpha_i},
\label{energy3}
\end{equation}
which can be evaluated, at least by means of some numerical
methods, for all values of the parameters $m_i, n_i, \alpha_i$. \\
One can notice that behavior of the integral (and difficulty of
the calculation) strongly depends on a particular value of the
following ration
\begin{equation}
\frac{n^2_i}{m^2_i} = q^2_i. \label{def_q}
\end{equation}
In the next section the simplest case, when all these numbers are
equal will be investigated.
%%%%%%%%%%%%%%%%%%%%%%%%%%%%%%%%%%%%%%%%%%
\section{\bf{ Case $q^2=const.$}}
%%%%%%%%%%%%%%%%%%%%%%%%%%%%%%%%%%%%%%%%%%
Let us focus on the case when
\begin{equation}
\frac{n^2_i}{m^2_i} = q^2=const., \label{cond3}
\end{equation}
for all $i=1...N$. Then the equations of motion (\ref{eq_mot6})
can be simplified to the following expressions
\begin{equation}
\frac{f_jf'_j}{(1+f_j^2)^2}= \frac{1}{k_j} \prod_{i=1}^N
k_i^{4\alpha_i} \left( m^2_i +\frac{n^2_i}{\sinh^2 \eta
}\right)^{-2\alpha_i} \frac{1}{\sinh^2 \eta}. \label{eq_mot7}
\end{equation}
Moreover, these equations can be integrated out and finally we
obtain that
\begin{equation}
\frac{1}{1+f^2_j}=\frac{1}{k_j} \frac{1}{1-q^2} \left(
\prod_{i=1}^N k_i^{4\alpha_i} m_i^{-4\alpha_i} \right) \frac{
\cosh \eta }{\left( q^2+\sinh^2 \eta \right)^{\frac{1}{2}} }+l_j,
\label{sol1}
\end{equation}
where $j=1...N$. Here we have taken $q^2<1$. In order to find the
value of the integration constants we need to specify the
asymptotical conditions for the functions $f_j$. We choice them in
the form which admits the nontrivial topological structure
\cite{aratyn}
\begin{equation}
\vec{n} \rightarrow (0,0,1) \; \; \mbox{i.e.} \; \; f \rightarrow
\infty \; \; \mbox{as} \; \; \eta \rightarrow 0 \label{bound1}
\end{equation}
and
\begin{equation}
\vec{n} \rightarrow (0,0,-1) \; \; \mbox{i.e.} \; \; f \rightarrow
0 \; \; \mbox{as} \; \; \eta \rightarrow \infty. \label{bound2}
\end{equation}
Then we find that
\begin{equation}
\frac{\prod_{i=1}^N k_i^{4\alpha_i}}{k_j}=\frac{1}{2 \prod_{i=1}^N
m^{-4\alpha_i} } \frac{1-q^2}{|q|-1} |q|. \label{stala_k}
\end{equation}
This is equivalent to the fact that all integration constants have
to take the same value
\begin{equation}
k_j=k=const. \label{cond_stala1}
\end{equation}
In addition, one can obtain that
\begin{equation}
l_j=l= -\frac{1}{ |q|-1}. \label{stala_l}
\end{equation}
Finally, the solution is given by the following formula
\begin{equation}
\frac{1}{1+f^2_j}=\frac{1}{1-|q|} \left[ 1-|q| \frac{\cosh
\eta}{q^2 + \sinh^2 \eta} \right]. \label{sol2}
\end{equation}
One can recognize here the set of $N$ Aratyn-Fereira-Zimerman
toroidal solutions. \\
Let us now show that the solution possesses toroidal symmetry. The
components of the unit vector fields read
\begin{equation}
n^1_{(i)}=\frac{2f_i}{1+f^2_i} \cos (m_i\xi +n_i\phi),
\label{sol_n1}
\end{equation}
\begin{equation}
n^2_{(i)}=\frac{2f_i}{1+f^2_i} \sin (m_i\xi +n_i\phi),
\label{sol_n2}
\end{equation}
\begin{equation}
n^3_{(i)}=\frac{-1+f^2_i}{1+f^2_i}, \label{sol_n3}
\end{equation}
where $f_i$ is given by (\ref{sol2}) i.e. depends only on the
radial-like coordinate $\eta$. It is clearly visible that surfaces
of constant $n^3_{(i)}$ are torii.
\\
After inserting obtained here solution into the total energy
integral one finds that
\begin{equation}
E_{m,n}=(2\pi)^2 4\cdot 2^{1/4} \prod_{i=1}^N m_i^{2\alpha_i}
\sqrt{1+|q|} \sqrt{|q|}. \label{energy_q1}
\end{equation}
In the limit $m_{\alpha_i}=m, \; i=1...N$ the formula founded in
\cite{aratyn} can be reproduced. \\
In order to prove that our solutions possess non-trivial topology
one has to calculate value of the pertinent topological invariant.
In our case static field configurations, which can be viewed as
maps from $S^3$ into $S^2$, are classified by so-called Hopf index
$Q_H=\pi_3 (S^2)$. One should notice that because of the fact that
we deal with $N$ fields thus $N$
different topological Hopf charges $Q_H^{(i)}$ have to be introduced. \\
We define two additional functions
\begin{equation}
g_1^2=   \cosh \eta - \sqrt{\frac{n^2}{m^2} +\sinh^2 \eta}
\label{g1}
\end{equation}
and
\begin{equation}
g_2^2=  \sqrt{1+\frac{m^2}{n^2}\sinh^2 \eta} -\cosh \eta .
\label{g2}
\end{equation}
Then, for every field we introduce four objects
\begin{equation}
\Phi_{\left( ^1 _2 \right)}^{(i)}=\left(
\frac{g_1}{\sqrt{g_1^2+g_2^2}}\right) \times \left( ^{\cos m_i\xi}
_{\sin m_i \xi}  \right) \label{topPhi1}
\end{equation}
and
\begin{equation}
\Phi_{\left( ^3 _4 \right)}^{(i)}=\left(
\frac{g_2}{\sqrt{g_1^2+g_2^2}}\right) \times \left( ^{\cos
n_i\phi} _{-\sin n_i \phi}  \right). \label{topPhi3}
\end{equation}
They are related with the i-th unit vector field via the standard
relations $\vec{n}_i=Z^{\dagger i} \vec{\sigma} Z^i$, where
$\vec{\sigma}$ are Pauli matrices and
\begin{equation}
Z^i=\left(
\begin{array}{c}
Z_1^i \\
Z_2^i
\end{array}
\right), \; \; \; Z^{i \dagger} =( Z_1^{i*}, Z_2^{i*})
\label{parametrHopf}
\end{equation}
and
\begin{equation}
Z_1^i=\Phi_1^{(i)}+i\Phi_2^{(i)}, \; \; \; Z_2^i=\Phi_3^{(i)}
+i\Phi_4^{(i)}. \label{parametrZ}
\end{equation}
Using these objects one can construct $N$ Abelian vector fields
which form the previously introduced antisymmetric field tensors
$H^{(i)}_{kl}=\partial_k A_l^{(i)} -\partial_l A_k^{(i)}$. In fact
one finds that
\begin{equation}
A_k^{(i)} = \frac{i}{2} (Z^{i \dagger} \partial_k Z^i -\partial_k
Z^{i \dagger} Z^i). \label{ab-potential}
\end{equation}
Then, using the standard expression for the Hopf index
\begin{equation}
Q_H^{(i)} =\frac{1}{4\pi^2} \int d^3x \vec{A}^{(i)} \cdot
\vec{B}^{(i)}, \label{indexHopf}
\end{equation}
we arrive at the following result
\begin{equation}
Q_H^{(i)}=\frac{n_im_i}{2}
\left[((\Phi^{(i)}_1)^2+(\Phi_2^{(i)})^2)^2 -
((\Phi_3^{(i)})^2+(\Phi_4^{(i)})^2)^2 \right]_0^{\infty} =-n_im_i.
\label{hopf}
\end{equation}
In other words our soliton solutions are classified by set of $N$
Hopf charges. Every kind of the unit vector field carries its own
topological charge which is unique determined by the parameters
$m_i, n_i$. Now, one could ask about generalization of
Vakulenko-Kapitansky inequality for $N$ vector fields. Knowing
that $\sqrt{1+|q|} \sqrt{q} \geq \sqrt{2} |q|^{3/4}$ we obtain the
generalized relation
\begin{equation}
E \geq (2\pi)^2 4 \cdot 2^{\frac{3}{4}} \prod_{i=1}^N
Q_{(i)}^{\alpha_i}, \label{gen_VK}
\end{equation}
where $\sum_{i=1}^N \alpha_i =\frac{3}{4}$. If all topological
numbers are equal $Q_{(i)}=Q$, $i=1..N$ we reproduce the standard
Vakulenko-Kapitansky formula.
%%%%%%%%%%%%%%%%%%%%%%%%%%%%%%%%%%%%%%%%%%%%%%%%%%%%%%%%%%%%%%%
\section{\bf{ Conclusions}}
%%%%%%%%%%%%%%%%%%%%%%%%%%%%%%%%%%%%%%%%%%%%%%%%%%%%%%%%%%%%%%%
In the present work a generalized Aratyn-Ferreira-Zimerman model
has been analyzed. This model consists of $N$ unit vector fields
and possesses $N$ independent $O(3)$ symmetries. We have proved
that this model is integrable in the sense that infinite number of
the conserved currents appears. Precisely speaking, one can find a
set of $N$
infinite families of the currents. \\
We have also shown that, as in the case of the soliton theories in
$(2+1)$ dimensions, this property leads to the existence of
non-trivial topological field configurations. Our soliton
solutions are described by shape functions depending on $\eta$
coordinate and angular terms characterized by set of two integer
parameters $m_i$ and $n_i$. The total energy has been also found.
In general, it is given by the integral (\ref{energy3}). The exact
form of the shape functions has been obtained in the limit of the
constant ratio of the Ansatz parameters $n^2_i/m^2_i =q^2=const.$
Then the shape functions take form of the well-know
Aratyn-Ferreira-Zimerman solution. However, obtained solutions are
much more general due to the angular part which for each of the
unit vector field can be different. Because of that the
topological behavior of the solutions is changed - now we have $N$
independent topological Hopf indexes. In other words we have
obtained multi-soliton configurations classified by $N$
topological numbers $Q_H^{(i)}$, where each of the vector field
carries different topological charge. These configurations consist
of $N$ toroidal solitons of the different types. \\
It should be stressed that all properties of the solutions
(topological charge, total energy), at least in the $q_i^2=const.$
case, are determined only by value of the parameters $m_i, n_i$.
For example, behavior of the shape functions $f_i$ depends only on
the ratio between them whereas number of the fields does not
affect the shape functions. Regardless of the particular number of
the fields $N$, all functions $f_i$ possess identical form. On the
other hand, number of the vector fields defines the topological
content of the model. Similarly, particular values of the
parameters $\alpha_i$ of the discussed model also do not inflect
on the shape of solitons as well as on their topological charges.
\\
In our opinion, the most interesting result is the generalized
Va\-ku\-len\-ko-Ka\-pi\-tan\-ski inequality. As we see this
formula can be divided into two parts. The first part contains
only numbers and seems to be universal i.e. Lagrangian
independent. In fact, it is identical as in the simple one vector
field case. The second part, which includes the Hopf indexes,
appears to be strongly sensitive to the form of the model. To show
it more precisely one can consider a simple sum of $N$ standard
Aratyn-Ferriera-Zimerman Lagrangians
$$\mathcal{L}=\sum_{i=1}^N ( H_{\mu \nu}^{(i) 2} )^{3/4}$$ i.e. a
model with the noninteracting unit vector fields. One can
immediately show that the pertinent Vakuleko-Kapitanski formula
reads $$E \geq (2\pi)^2 4 \cdot 2^{\frac{3}{4}} \sum_{i=1}^N
|Q^{(i)}|^{\frac{3}{4}}.$$ We see that the second part of the
inequality reflects a form of the Lagrangian. However, a common
feature is shared by both inequalities - there is an universal
scaling property. Namely, if we transform all topological charges
$Q^{(i)} \rightarrow \lambda Q^{(i)}$ then the total energy will
scale as $E \rightarrow
\lambda^{\frac{3}{4}} E$. \\
Let us now analyze the physical meaning of the parameters
$\alpha_i$. In spite of the fact that the form of the solution, at
least in the $q_i^2=const.$ regime, does not depend on their
values, they play crucial role in the interaction of the toroidal
solitons. Indeed, interaction between hopfions in the proposed
model is governed by the values of $\alpha_i$ and it can be
attractive or repulsive. Quite interesting hopfions can even do
not interact at all. To be more precisely we will consider two
hopfions with the following topological charges
$Q_a=(Q_a^{(1)},..,Q_a^{(N)})$, $a=1,2$. It follows from the
formula (\ref{gen_VK}) that if $\alpha_i<1$ then $i$-th hopfions
built of the $i$-th vector field should attract each other whereas
for $\alpha_i >1$ this interaction will be repulsive. In the very
special case where $\alpha_i=1$, they do not interact at all. It
resembles the Bogomolny limit in the standard soliton systems. As
there is only one constrain on the parameters $\alpha_i$
(\ref{stab_cond}), all three situations can be realized. Thus, in
general, one can deal with a model where every possible type of
interaction occurs.
\\
As we have mentioned it above, the analytical solutions have been
obtained only in the simplest situation i. e. for $q_i^2 =const.$.
It constrains also the validity of the generalized
Vakulenko-Kapitanski inequality. Thus one should investigate more
complicated cases with $q_i^2 \neq q_j^2$ as well. We would like
to address this issue in our next paper. \vspace{5mm}

\noindent This work is partially supported by Foundation for
Polish Science and ESF "COSLAB" programme.

\end{document}